\documentclass[12pt]{article}

\usepackage[dvips]{epsfig}
\usepackage{amsfonts}

\usepackage{amsmath}
\usepackage{amssymb}
\allowdisplaybreaks

\begin{document}

\baselineskip=16.8pt plus 0.2pt minus 0.1pt


\makeatletter
\@addtoreset{equation}{section}
\renewcommand{\theequation}{\thesection.\arabic{equation}}
\renewcommand{\thefootnote}{\fnsymbol{footnote}}

\newcommand{\ap}{\alpha '}
\newcommand{\p}{\partial}
\newcommand{\be}{\begin{equation}}
\newcommand{\ee}{\end{equation}}
\newcommand{\nn}{\nonumber}
\newcommand{\ds}{\displaystyle}
\newcommand{\wt}[1]{\widetilde{#1}}
\newcommand{\bm}[1]{\boldsymbol{#1}}
\newcommand{\ket}[1]{\vert #1\rangle}
\newcommand{\abs}[1]{\left\vert #1\right\vert}
\newcommand{\QB}{Q_{\rm B}}
\newcommand{\go}{g_{\rm o}}
\newcommand{\toc}{t_1^{\,\rm c}}
\newcommand{\subt}{\widehat{t}}
\newcommand{\calO}{{\cal O}}

\begin{titlepage}
\title{
\hfill\parbox{4cm}
{\normalsize KUNS-1835\\{\tt hep-th/0304163}}\\
\vspace{1cm}
{\bf Time Dependent Solution in Cubic String Field Theory
}
}
\author{
Masako {\sc Fujita}\thanks{
{\tt masako@gauge.scphys.kyoto-u.ac.jp}}
\
and
\
Hiroyuki {\sc Hata}\thanks{
{\tt hata@gauge.scphys.kyoto-u.ac.jp}}
\\[15pt]
{\it Department of Physics, Kyoto University, Kyoto 606-8502, Japan}
}
\date{\normalsize April, 2003}
\maketitle
\thispagestyle{empty}
\begin{abstract}
\normalsize

We study time dependent solutions in cubic open string field theory
which are expected to describe the configuration of the rolling
tachyon.
We consider the truncated system consisting of component fields
of level zero and two, which are expanded in terms of $\cosh n x^0$
modes.
For studying the large time behavior of the solution we need to know
the coefficients of all and, in particular, large $n$ modes.
We examine numerically the coefficients of the $n$-th mode, and find
that it has the leading $n$-dependence of the form
$(-\beta)^n\,\lambda^{-n^2}$ multiplied by a peculiar subleading part
with peaks at $n=2^m=4,8,16,32,64,128,\cdots$.
This behavior is also reproduced analytically
by solving simplified equations of motion of the tachyon system.

\end{abstract}
\end{titlepage}

\section{Introduction}

Recently the time dependent decay of an unstable D-brane has been
one of the most active subjects in string theory.
The unstable D-brane evolves in time
as the tachyon field rolls down to the bottom of the potential
corresponding to the stable closed string vacuum.
Sen proposed a boundary conformal field theory (BCFT)
which describes this time dependent process \cite{rolling}.
This BCFT is exactly solvable \cite{Callan,Polchinski,Recknagel}
and he found that in the limit of vanishing string coupling
the system evolves to a pressureless gas called tachyon
matter \cite{tmatter}.
Since tachyon matter is classically stable, it could be applied to
cosmology as inflaton, and the tachyon matter cosmology has attracted
great interest.

Besides such cosmological applications, time dependent solutions
themselves are of importance in string theory.
For they are expected to reveal new nonperturbative aspects of string
dynamics.
Much work has been done for the study of the rolling tachyon
solutions using various formulations such as conformal field theory,
effective field theory, boundary string field theory,
supergravity, and so on
\cite{OkudaSugimoto,evolution,closed,nofreedom,TimeTachyon,hashimoto,
GHY,minahan,SugimotoTerashima,yokono,Chen,Maldacena}.
However, there are many problems left unresolved.
For example, closed string emission and its back reaction in the
process of rolling \cite{OkudaSugimoto,Chen,Maldacena}.

In this paper, we study time dependent solutions describing the
rolling tachyon in cubic string field theory (CSFT) \cite{Witten-SFT}.
We construct time dependent solutions in truncated CSFT
and examine whether the solutions have the properties of
the rolling tachyon.
The desirable properties are as
follows:
\begin{enumerate}
\item
There is a one-parameter family of solutions.
This parameter $\beta$ corresponds to the initial value
of tachyon field.
\item
For $\beta =0$, the tachyon stays at the unstable perturbative vacuum:
\be
t(x;\beta=0 )=0.
\ee
\item
There is a specific value $\beta_c$ of $\beta$.
At $\beta =\beta_c$ the tachyon field is
independent of time and stays at the tachyon vacuum:
\be
t(x^0;\beta_c)=t_c.
\ee
\item
For $0<\beta <\beta_c$, the tachyon evolves in time and
asymptotically approaches the tachyon vacuum $t_c$:
\be
t(x\rightarrow\infty ;\beta\neq 0 )=t_c.
\ee
\end{enumerate}

We expand the component fields in
terms of $\cosh nx^0$ (and $\sinh nx^0$) and solve the equations of
motion for the modes \cite{marginal,padic}.
For example, in the truncated system of only the level zero tachyon
field $t(x^0)$, we express it as
\be
t(x^0)=\sum_{n=0}^{\infty}t_n\cosh nx^0,
\label{t=sum}
\ee
and solve the equations of motion for $t_n$ ($n\ne 1$) by taking $t_1$
as a given parameter corresponding to the marginal deformation
parameter in the full CSFT system.\footnote{
See \cite{padic,Aref'eva,Volovich} for approaches to solving the
differential equations of the component fields without using the
mode expansion.
}
However, if we truncate the summation in \eqref{t=sum} at $n=N$ as in
the modified level truncation scheme \cite{lump2}, the solution
diverges like $\cosh Nx^0$ as $x^0\rightarrow\infty$.
A hint on avoiding this divergent behavior is found in the calculation
of the coefficient $f(x^0)$ of the closed string tachyon in the
rolling tachyon boundary state \cite{rolling}.
There, an infinite summation of $\cosh nx^0$  gives a finite value at
$x^0\rightarrow\infty$ by analytic continuation:
\be
f(x^0)=1+2\sum_{n=1}^{\infty}(-\wt{\beta}\,)^n \cosh nx^0=
\frac{1}{1+\wt{\beta}\,e^{x^0}}+\frac{1}{1+\wt{\beta}\,e^{-x^0}}-1 .
\ee
Moreover, at a specific value of the parameter $\wt{\beta}$;
$\wt{\beta}=1$, the summation becomes a constant,
which is a desirable property for the rolling tachyon solution.
This example suggests that, for getting a convergent function at
$x^0\to\infty$, we have to obtain the coefficients $t_n$
for all, and in particular, large $n$ and carry out the infinite
summation of \eqref{t=sum}.

In this paper, we first solve numerically the equation of motion of
$t_n$ to study its  $n$-dependence. We find that $t_n$ has a leading
$n$-dependence of the form $(-\beta)^n\lambda^{-n^2}$ with
$\beta$ being proportional to the parameter $t_1$. Besides this leading
$n$-dependence, $t_n$ has a peculiar subleading part with peaks at
$n=2^m=4,8,16,32,64,128,\cdots$.
Then we carry out analytical study of the equation of motion of $t_n$.
Since the original equation of motion is too complicated to be solved
analytically, we consider an approximate equation of motion valid for
a large $n$. This equation is further deformed to a simplified one
which still maintains the essential features of the original equation.
In fact, this simplified equation can be solved analytically at
special values of $n$ to reproduce the behavior of $t_n$ found by
numerical analysis.

The rest of this paper is organized as follows.
In sec.\ 2, we present the CSFT action and the equation of motion
of $t_n$. In sec.\ 3, we carry out the numerical analysis of $t_n$ and
also the level two modes, and find their $n$-dependence.
Then in sec.\ 4, the equation of motion of $t_n$ is examined
analytically to reproduce their $n$-dependence found in sec.\ 3.
The final section (sec.\ 5) is devoted to a summary and discussions.
In the appendix, we present technical details used in sec.\ 4.

\section{CSFT action and the equation of motion}

The action of cubic string field theory for bosonic open string
takes the form \cite{Witten-SFT}
\be
S=-\frac{1}{\go^2}\left(
\frac12
\Phi\cdot\QB\Phi + \frac13 \Phi\cdot\left(\Phi\ast\Phi\right)\right),
\label{csft}
\ee
where $\QB$ is the BRST operator and $\ast$ is the star product
between two string fields.
The open string field $\Phi$ contains component fields
corresponding to all the states in the first-quantized string Fock space.
For more details of CSFT and its application to tachyon condensation,
see \cite{ohmori} and the references therein.
First let us keep only the tachyon field $t(x)$ in $\Phi$,
$\ket{\Phi}=b_0 \ket{0}t(x)$.
The truncated action including only the tachyon field is
\be
S=\frac{1}{\go^2}\int d^{26}x\left(
\frac12\,  t(x)\,(\Box+1)\, t(x)
-\frac13 \lambda\left(\lambda^{(1/3)\Box}t(x)\right)^3\right),
\label{taction}
\ee
with
\be
\Box=-\p_0^2+\nabla^2 ,\qquad \lambda=3^{9/2}/2^6=2.192 .
\ee
We use the convention of $\ap=1$ throughout this paper.
Later we shall take into account of the level two fields.

In the following we are interested in the time dependent and
spatially homogeneous solution $t(x^0)$ to the equation of motion
derived from (\ref{taction}):
\be
(\p_0^2-1)t(x^0)+\lambda^{1-\p_0^2/3}
\left(\lambda^{-\p_0^2/3}t(x^0)\right)^2=0.
\label{eq_t}
\ee
As we mentioned in sec.\ 1,
we expand the tachyon field in terms of $\cosh nx^0$
following \cite{padic,marginal}.
Here, we do not adopt the modified level truncation scheme
\cite{lump2}, but take into account of all the modes:
\be
t(x^0)=\sum_{n=0}^{\infty}t_n \cosh nx^0 .
\label{expand}
\ee
Substituting this into the equation of motion \eqref{eq_t},
we have the following equations for each $t_n$:
\begin{align}
&-t_0 +\lambda\left[
(t_0)^2 +\frac12\sum_{n=1}^\infty \lambda^{-2 n^2/3}\,(t_n)^2
\right]=0,
\label{eq_t0}
\\
&2\, t_0\, t_1
+\sum_{k=1}^{\infty}
\lambda^{-2k(k+1)/3}\, t_k\, t_{k+1}=0,
\label{eq_t1}
\\
&(n^2-1)\,t_n  +\lambda^{1-2n^2/3}\biggr[2\, t_0\, t_n
+\frac12\sum_{k=1}^{n-1}\lambda^{-2k(k-n)/3}\, t_k\, t_{n-k}
\nn\\
&\hspace{6cm}
+\sum_{k=1}^{\infty}
\lambda^{-2k(k+n)/3}\, t_k\, t_{n+k}\biggr]=0\quad (n\ge 2),
\label{eq_tn}
\end{align}
which are respectively the equations of motion of $t_0$, $t_1$ and
$t_n$ ($n\ge 2$).

{}From the BCFT analysis of rolling tachyon \cite{rolling},
it is expected that the full CSFT keeping all the component fields in
$\Phi$ has a rolling solution with one free parameter
corresponding to the initial value of the tachyon field at $x^0=0$ .
However, in the approximate analysis with finite number of component
fields, we no longer have such a free parameter.
Therefore, in our analysis we treat $t_1$ as a given parameter and
solve \eqref{eq_t0} and \eqref{eq_tn} for
$t_n(t_1)$ ($n=0,2,3,\cdots$) \cite{padic,marginal}.
In the exact CSFT, we can in principle adopt any $t_n$ as the free
parameter.
However, in our approximate analysis the result depends on choice of
the free parameter.
The present choice is natural since,
in the case without the interaction in \eqref{eq_t}, the solution is
given by $t_1\cosh x^0$ with $t_1$ being arbitrary.

Before starting the numerical analysis of $t_n(t_1)$, we shall mention
that the parameter $t_1$ has an upper bound $\toc$ above which
\eqref{eq_t0} and \eqref{eq_tn} have no real solutions \cite{marginal}.
To understand this fact it is sufficient to consider only $t_0$
neglecting other $t_n$ with $n\ge 2$, since they are negligibly
small compared with $t_0$ as we shall see in later sections.
Solving \eqref{eq_t0} with $t_n=0$ ($n\ge 2$), we have two branches of
solutions \cite{marginal},
\be
t_0^{(\pm)}(t_1)=\frac{1}{2\lambda}\!\left(1\pm\sqrt{1-
\left(\frac{t_1}{\toc}\right)^2}\right) ,
\label{t0pm}
\ee
with $\toc$ given by
\be
\toc=\frac{1}{\sqrt{2}}\,\lambda^{-2/3}
=0.419 .
\label{deftoc}
\ee
The solutions $t_0^{(+)}$ and $t_0^{(-)}$ are those connected with the
tachyon vacuum $t(x)=1/\lambda$ and the unstable perturbative vacuum
$t(x)=0$, respectively, and $t_0^{(-)}$ is the rolling solution we are
interested in.
Both $t_0^{(\pm)}$ are real only in the region $\abs{t_1}\le\toc$.

\section{Numerical analysis}

In this section we shall carry out numerical analysis of the rolling
solutions in CSFT. In sec.\ 3.1, we solve numerically the equation of
motion of only the level zero tachyon field $t(x^0)$.
Then in sec.\ 3.2 we extend our analysis to the system with level two
fields.

\subsection{Numerical analysis for tachyon field}

\begin{figure}[htbp]
\begin{center}
\leavevmode
\put(0,190){\boldmath{$-\ln\abs{t_n}$}}
\put(232,56){\boldmath{$n$}}
\epsfig{file=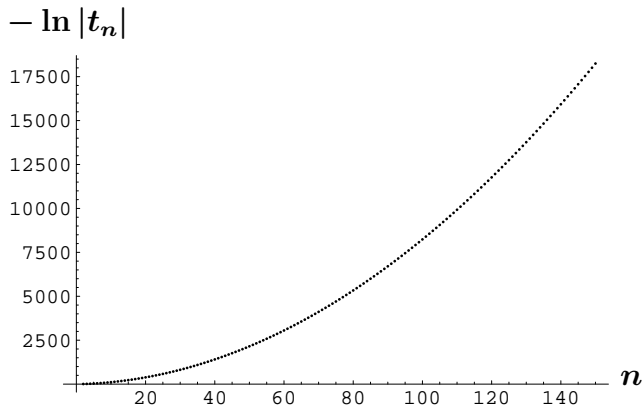, width=8cm}
\vspace{-1.0cm}
\caption{$-\ln\!\abs{t_n}$ for $n=2,3,\cdots,150$ at $t_1=0.1$.}
\label{lntn}
\end{center}
\end{figure}

As we mentioned in the previous section, we solve numerically the
equations of motion \eqref{eq_t0} and \eqref{eq_tn} for
$t_0$ and $t_n$ ($n\ge 2$) for a given value of $t_1$.
Since numerical calculation cannot be done for an infinite number of
variables $t_n$, we put $t_n\equiv 0$ for $n> N$ with sufficiently
large $N$ and solve \eqref{eq_t0} and \eqref{eq_tn} for $n\le N$.
Here, we take $N=150$ and obtain the solution $t_n$
($n=0,2,3,\cdots,150$) by Newton's method starting with the point
$t_n=0$
(namely, we are considering the solution which is reduced to the
perturbative vacuum in the limit $t_1\to 0$).
We have obtained $t_n(t_1)$ at discrete points
$t_1=0.01,0.02,0.03,\cdots,0.41$ and $0.418$.
There exists no real solutions for $t_1\gtrsim 0.419$, which is
consistent with the simplified analysis of \eqref{t0pm}.
Fig.\ \ref{lntn} shows $-\ln\abs{t_n}$ ($n\ge 2$) at $t_1=0.1$.

\begin{figure}[htbp]
\begin{center}
\leavevmode
\put(25,180){\boldmath{$\xi-1$}}
\put(232,91){\boldmath{$t_1$}}
\epsfig{file=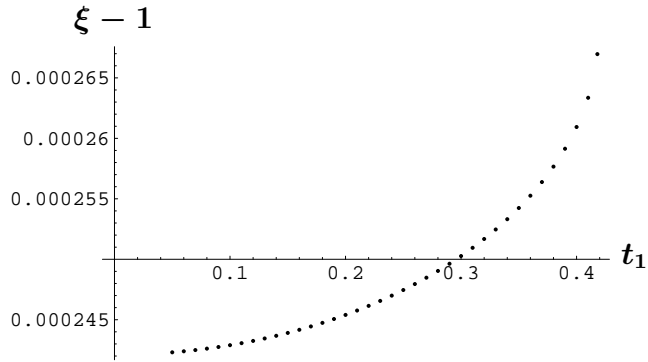, width=8cm}
\vspace{-1cm}
\caption{$\xi(t_1)-1$ at $t_1=0.05,0.06,\cdots,0.41$ and $0.418$.
The dots representing $\xi(t_1)$ for $t_1\leq 0.04$ are missing
since $t_n$ with larger $n$ vanishes in our numerical precision, which
makes it impossible to determine $\xi$ by fitting.}
\label{xit150}
\end{center}
\end{figure}

\begin{figure}
\begin{center}
\leavevmode
\put(17,170){\boldmath{$\beta$}}
\put(205,47){\boldmath{$t_1$}}
\epsfig{file=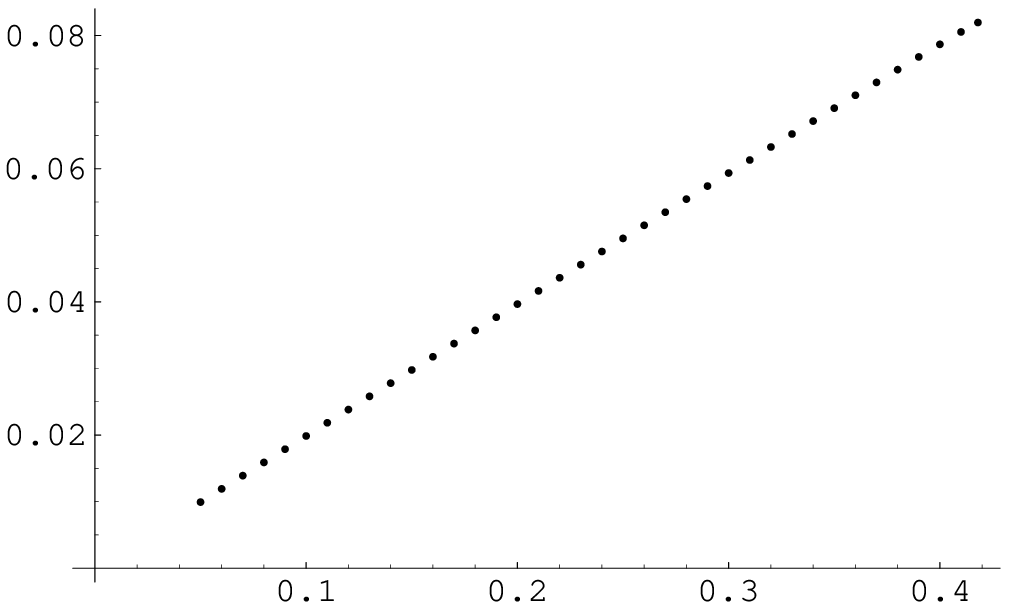, width=7cm}
\vspace{-1cm}
\caption{$\beta(t_1)$ at $t_1=0.05,0.06,\cdots,0.41$ and $0.418$.
For the same reason as in the case of $\xi(t_1)$ (fig.\ \ref{xit150}),
the dots are missing for $t_1\leq 0.04$.}
\label{betat150}
\end{center}
\end{figure}

\begin{figure}[htbp]
\begin{center}
\leavevmode
\put(10,168){\boldmath{$A$}}
\put(207,55){\boldmath{$t_1$}}
\epsfig{file=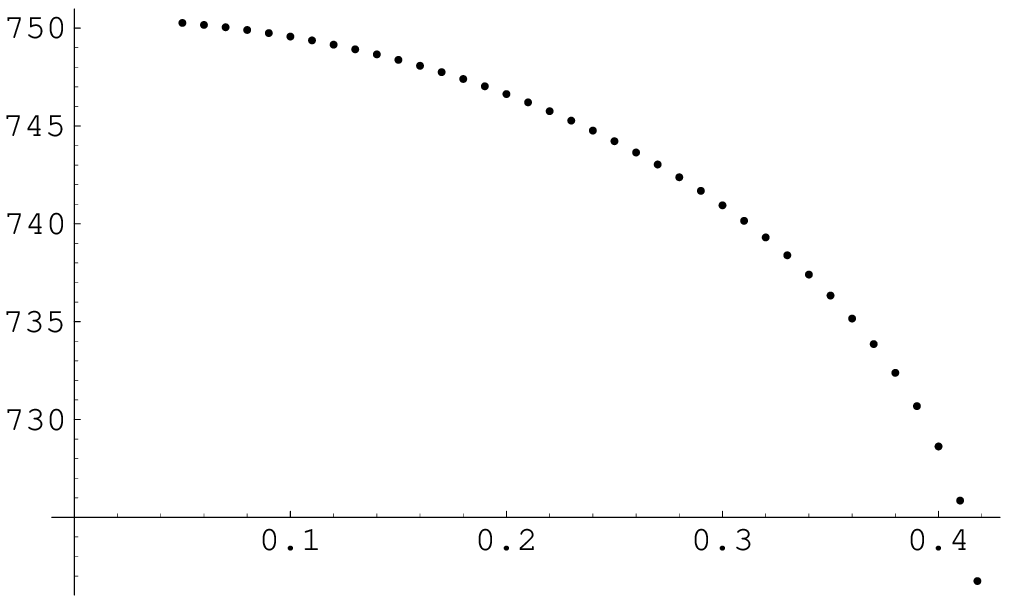, width=7cm}
\vspace{-1cm}
\caption{$A(t_1)$ at $t_1=0.05,0.06,\cdots,0.41$ and $0.418$.
For the same reason as in the case of $\xi(t_1)$ (fig.\ \ref{xit150}),
the dots are missing for $t_1\leq 0.04$.}
\label{constant}
\end{center}
\end{figure}

Then, let us assume that the $n$-dependence of $t_n$ at each value of
$t_1$ is given by
\be
t_n^{\rm fit}=-A\,\lambda^{-\xi n^2}(-\beta)^n ,
\qquad(n\geq 2),
\label{tnfit}
\ee
and obtain $A(t_1)$, $\xi(t_1)$ and $\beta(t_1)$ by fitting to our
numerical solution $t_n(t_1)$.
This assumption of \eqref{tnfit} will be justified by analytical
considerations in the next section.
\begin{figure}[htbp]
\begin{center}
\leavevmode
\put(0,190){\boldmath{$-\ln\abs{t_n}$}}
\put(239,55){\boldmath{$n$}}
\vspace{-1cm}
\epsfig{file=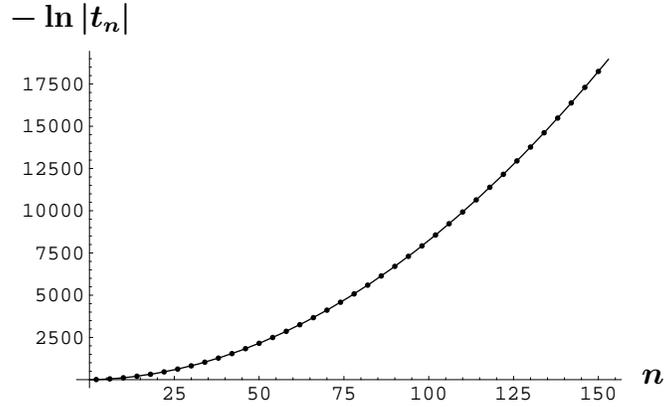, width=8cm}
\vspace{-0.5cm}
\caption{The numerical solution $-\ln\abs{t_n}$ (dots) and the
fitted curve
$0.7851\, n^2+ 3.929\, n  -6.620 $, which corresponds to
$\xi =1.00024$, $\beta=0.01986$ and $A=749.6$.
For the sake of visibility, the dots representing the numerical
solution are plotted only for $n=2,6,10,14,\cdots,150$.
}
\label{fitt150}
\end{center}
\end{figure}
The solution $t_n$ and the fitted curve \eqref{tnfit} at, for example,
$t_1=0.1$ are shown in fig.\ \ref{fitt150}, which confirms the validity
of our assumption \eqref{tnfit}.
Fitting by \eqref{tnfit} works very well also at other values of $t_1$.
Figs.\ \ref{xit150}, \ref{betat150} and \ref{constant} show
$\xi(t_1)$, $\beta(t_1)$ and $A(t_1)$ obtained by fitting.
These results show that
\begin{align}
&\xi(t_1)\simeq 1 ,
\label{xisimeq1}
\\
&\beta(t_1)\propto t_1\ \left(\simeq 0.196\, t_1\right) ,
\label{betaproptot1}
\end{align}
and that $A(t_1)$ is almost a constant.
Fig.\ \ref{xit150} (with $N=150$) itself tells that $\xi\simeq 1.00026$
with little $t_1$-dependence. For other values of the cutoff $N$, we
have $\xi\simeq 1.0045$ ($N=50$) and $\xi\simeq 1.00148$ ($N=100$).
These results strongly suggest that $\xi(t_1)$ approaches a constant
which is equal to one in the limit $N\to\infty$.
As for $\beta(t_1)$, the proportionality constant ($0.196$ for $N=150$)
for other values of $N$ are $0.2391$ ($N=50$) and $0.2137$ ($N=100$).
The proportionality constant gradually decreases as $N$ becomes
larger, and the value at $N=\infty$ predicted by the fitting using
$a +b/N$ is $0.18$.
The value of $\beta$ at $t_1=0.418$ which is close to the critical
$t_1$ above which we have no real solutions is
$\beta(0.418)=0.0819$ for $N=150$.
For other values of $N$, we have $\beta(0.418)=0.0999$ ($N=50$) and
$\beta(0.418)=0.0893$ ($N=100$).

\begin{figure}[htbp]
\begin{center}
\leavevmode
\put(10,180){\boldmath{$\subt_n$}}
\put(220,50){\boldmath{$n$}}
\put(245,180){\boldmath{$\subt_n$}}
\put(455,50){\boldmath{$n$}}
\epsfig{file=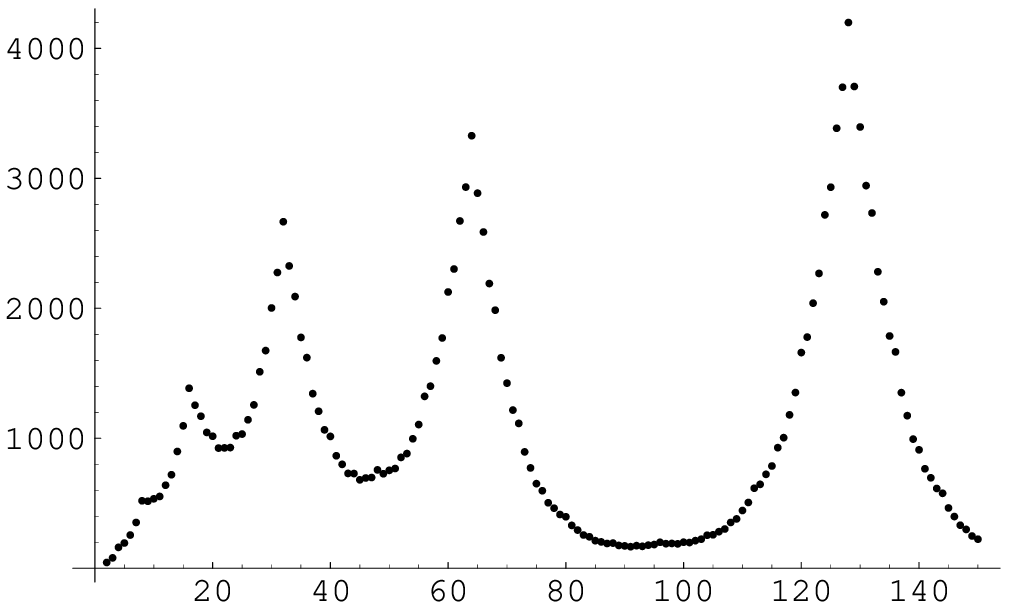, width=7.5cm}
\hspace{5mm}
\epsfig{file=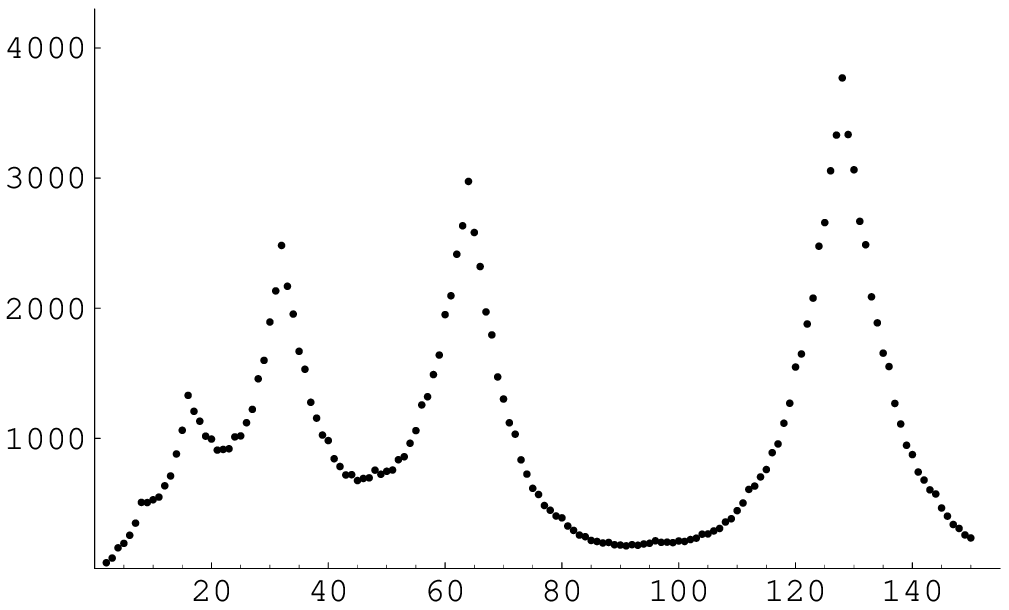, width=7.5cm}
\vspace{-1cm}
\caption{$\subt_n$ at $t_1=0.1$ (left figure) and $t_1=0.4$
(right figure). There are peaks at $n=4,8,16,32,64,128$. The shapes
of these two figures are almost the same. }
\label{subleading}
\end{center}
\end{figure}

Then let us consider the deviation of $t_n$ from the leading
$n$-dependence \eqref{tnfit}. Fig.\ \ref{subleading} shows
\be
\subt_n=-\frac{t_n}{\lambda^{-\xi n^2}(-\beta)^{n}},
\label{subt_n}
\ee
at $t_1=0.1$ and $0.4$
($\xi$ and $\beta$ in the denominator are those given by figs.\
\ref{xit150} and \ref{betat150}).
The coefficient $A$ of \eqref{tnfit} is the average value
of $\subt_n$ with respect to $n$.
As seen from fig.\ \ref{subleading}, the subleading part $\subt_n$ has
a peculiar $n$-dependence; it has peaks at $n=2^m=4,8,16,32,64,128$.
In addition, $\subt_n(t_1)$ depends very little on the value of $t_1$.
These properties of $\subt_n$ will be partially reproduced
by analytical consideration in sec.\ 4.

\begin{figure}[htbp]
\begin{center}
\leavevmode
\put(0,215){\boldmath{eq.\,\eqref{eq_t1}}}
\put(263,61){\boldmath{$t_1$}}
\epsfig{file=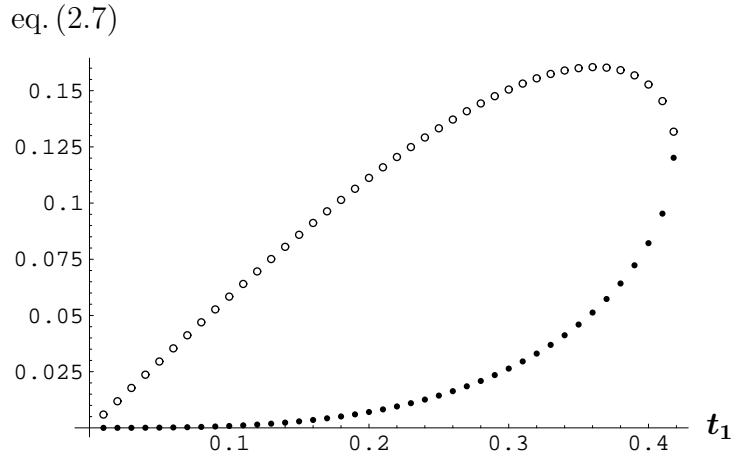, width=9cm}
\vspace{-1.5cm}
\caption{The LHS of \eqref{eq_t1} for the solution $t_n(t_1)$ obtained
by starting from the perturbative vacuum (dots) and that from the
tachyon vacuum (circles).}
\label{dV/dt}
\end{center}
\end{figure}

Finally, fig.\ \ref{dV/dt} shows the LHS of \eqref{eq_t1}
for the present solution $t_n(t_1)$ and the solution obtained by
starting from the tachyon vacuum.
It shows how well the equation of motion of $t_1$, which we do
not take into account in obtaining the solution, is satisfied.
As seen from the figure, the equation of motion of $t_1$ is fairly
well satisfied by the solution connected to the perturbative vacuum.
Since $t_n$ with larger $n$ are negligibly small as seen from
\eqref{tnfit}, the behavior of \eqref{eq_t1}
given by fig.\ \ref{dV/dt} is almost the same as that obtained by
the simplified analysis using \eqref{t0pm} and neglecting other $t_n$.

\subsection{Inclusion of level 2 fields}

We have carried out the same kind of analysis as in the previous
subsection by including the level two fields.
Taking into account of spatial rotational symmetry,
twist symmetry and the invariance under $X\to -X$ \cite{lump2},
the string field $\ket{\Phi}=b_0\ket{\phi}$ in the Siegel gauge
with component fields up to level two is expanded as
follows:
\begin{align}
\ket{\phi(x^0)}=& \ket{0}\, t(x^0)+c_{-1}b_{-1}\ket{0}\, u(x^0)
-\frac12 \left(\alpha_{-1}^0\right)^2 \ket{0}\, v(x^0)\nn\\
&\qquad\qquad\qquad - i\,\alpha_{-2}^0\ket{0}\,\chi(x^0)
+\frac12\alpha_{-1}^i\alpha_{-1}^i\ket{0}\, w(x^0),
\end{align}
where $t$, $u$, $v$ and $w$ are even functions of $x^0$ and hence
are expanded by $\cosh nx^0$; $\varphi(x^0)=\sum_{n=0}^\infty
\varphi_n \cosh nx^0$ ($\varphi=t$, $u$, $v$, $w$), while $\chi$ is an
odd function and is expanded by $\sinh nx^0$;
$\chi(x^0)=\sum_{n=1}^\infty\chi_n\sinh nx^0$.
These five fields are all hermitian.
We have used the level $(2,6)$ action
in our calculation.\footnote{
Here, we do not adopt the modified level truncation scheme as in
\cite{padic}, but we assign level zero to $t_n$ and level two to
$u_n$, $v_n$, $\chi_n$ and $w_n$ for all $n$.
The set of component fields in our level $(2,6)$ analysis is not
included nor includes that of the modified level $(4,8)$ analysis in
sec.\ 7 of \cite{padic}.
For the common component fields, the correspondence between
\cite{padic} (denoted by the superscript MZ) and the present paper
is 
$v_1^{\rm MZ} =(1/3)\left( 2\sqrt{2}\chi_1-v_1\right)$ and
$z_1^{\rm MZ} =(1/3)\left( v_1 - \chi_1/\sqrt{2}\right)$
($t_0$, $t_1$, $u_0$, $u_1$, $v_0$, $w_0$ and $w_1$ are the same
between the two).
Comparing the numerical solution at $t_1=0.05$ given in sec.\ 7 of 
\cite{padic} with ours, we find that $t_0$, $u_0$, $v_0$, $w_0$ and
$t_2$ agree within the accuracy of $0.1\%$, while
the agreement is only up to factor for $u_1$, $v_1$ and $w_1$.
}

\begin{figure}[htbp]
\begin{minipage}{0.46\hsize}
\begin{center}
\leavevmode
\put(20,170){\boldmath{$\xi$}}
\put(205,55){\boldmath{$t_1$}}
\epsfig{file=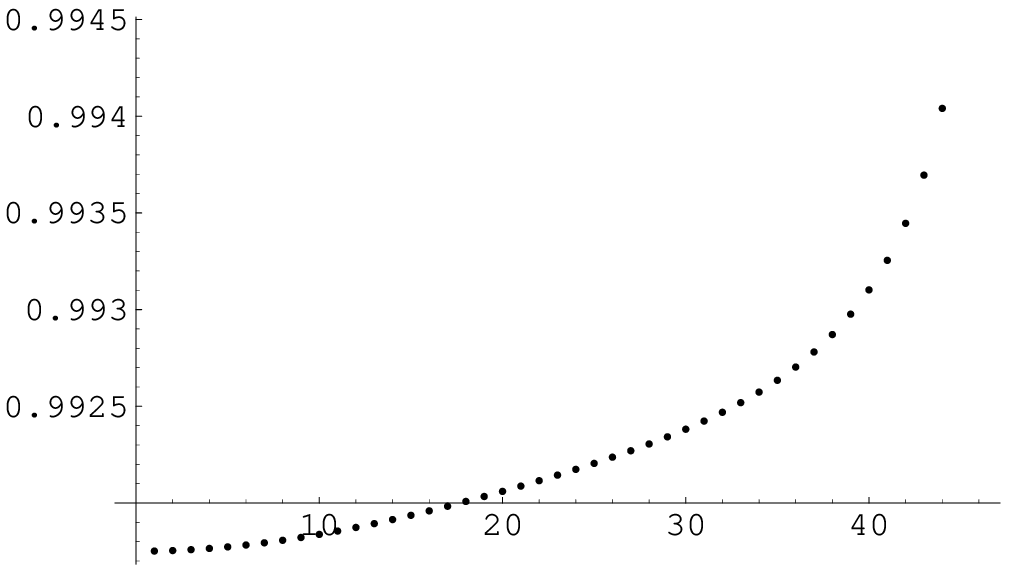, width=7cm}
\vspace{-1cm}
\caption{$\xi(t_1)$ at $t_1=0.01,0.02,\cdots,0.46$ in the level
  $(2,6)$ analysis. }
\label{xi3030}
\end{center}
\end{minipage}
\hspace{0.02\hsize}
\begin{minipage}{0.46\hsize}
\begin{center}
\leavevmode
\put(15,165){\boldmath{$\beta$}}
\put(205,47){\boldmath{$t_1$}}
\epsfig{file=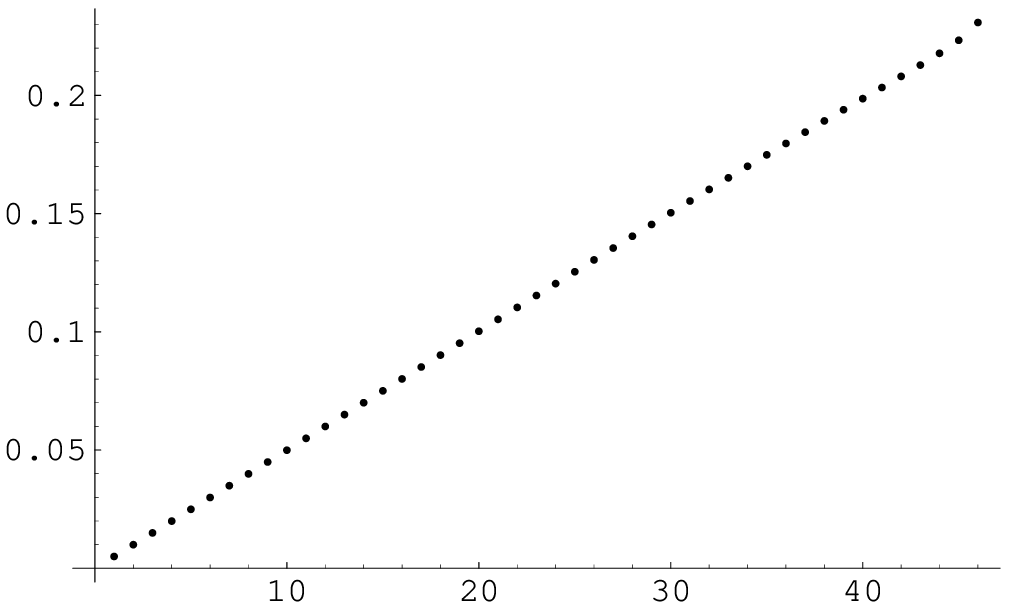, width=7cm}
\vspace{-1cm}
\caption{$\beta(t_1)$ at $t_1=0.01,0.02,\cdots,0.46$ in the level
  $(2,6)$ analysis.  The value of $\beta$ at $t_1=0.46$ is
  $\beta(0.46)=0.23$.}
\label{beta3030}
\end{center}
\end{minipage}
\end{figure}

In this numerical analysis, we adopted $N=30$ as the cutoff of the
mode number $n$ for all the five fields and solved the equation of
motion of each mode for a given value of $t_1$.
Then, fitting the solution $t_n(t_1)$ by \eqref{tnfit}, we obtained
$\xi(t_1)$ and $\beta(t_1)$, which are shown in figs.\ \ref{xi3030}
and \ref{beta3030}.\footnote{
In obtaining $\xi(t_1)$ and $\beta(t_1)$ by fitting, we omitted
$t_5$, $t_6$ and $t_{18}$.
Inclusion of them in the fit leads to
$\xi(t_1)$ and $\beta(t_1)$ with strange $t_1$-dependence
since $t_5$, $t_6$ and $t_{18}$ change their signs at $t_1\sim 0.30$,
$0.26$ and $0.37$, respectively.
}
The critical value of $t_1$ above which there is no real solution
is $\toc\simeq 0.46$ in the present case, which is larger than that in
the purely $t$ system.
Fig.\ \ref{xi3030} is consistent with our previous result
\eqref{xisimeq1}, $\xi(t_1)\simeq 1$.
{}From fig.\ \ref{beta3030} we have
\be
\beta(t_1)\simeq 0.50\, t_1.
\ee
In particular, the value of $\beta$ at the critical $t_1$ is
$\beta (t_1=0.46)\simeq 0.23$, which is about three times larger than
that in the purely $t$ system of sec.\ 3.1.
The $n$-dependences of other modes $\varphi_n$ ($\varphi=u,v,w,\chi$)
are similar to that of $t_n$.

In our level $(2,6)$ analysis with $N=30$,
the subleading part $\subt_n$ \eqref{subt_n} does not show such
characteristic behavior as in fig.\ \ref{subleading}.
This could be due to that $N=30$ is not sufficiently large.

\section{Analytical consideration}

In this section, we would like to understand analytically the
$n$-dependence of $t_n$ found in the previous section,
namely, \eqref{xisimeq1}, \eqref{betaproptot1} and the peculiar
behavior of the subleading part $\subt_n$ given
by fig.\ \ref{subleading}.

Let us consider the equation of motion \eqref{eq_tn} for large $n$.
We assume that the solution $t_n$ is given by
\be
t_n=-(-1)^n\frac{2}{\lambda}\, g_n\,\lambda^{-(2/3)cn^2}\qquad
(n\geq 1),
\label{ansatz}
\ee
where $c$ is a constant and $g_n$ has a weaker $n$-dependence than
the leading term $\lambda^{-(2/3)cn^2}$.
Concretely, we expect that $\ln g_n=\calO(n)$ for $n\gg 1$.
The front factor $(-1)^n 2/\lambda$ is simply for the sake of later
convenience. Note that, if $t_n$ is a solution, so is $(-1)^n t_n$.

We rewrite \eqref{eq_tn} into
\be
(n^2-1)t_n+\lambda^{1-(2/3)n^2}\left(2 t_0t_n+Y_n+Z_n\right)=0,
\label{eq_tn2}
\ee
where $Y_n$ and $Z_n$ are defined by
\begin{align}
Y_n&=\frac12\sum_{k=1}^{n-1}\lambda^{2k(n-k)/3}t_{n-k}t_k ,\\
Z_n&=\sum_{k=1}^{\infty}\lambda^{-2k(k+n)/3}t_{n+k}t_k .
\end{align}
Substituting the ansatz \eqref{ansatz} to $Y_n$ and $Z_n$,
we obtain
\begin{align}
Y_n&=\frac{2}{\lambda^2}(-1)^n\lambda^{-(2/3)(c/2 -1/4)n^2}
\sum_{k=1-n/2}^{n/2-1}\lambda^{-(2/3)(1+2c)k^2}
g_{\frac{n}{2}-k}g_{\frac{n}{2}+k},
\\
Z_n&=\frac{4}{\lambda^2}(-1)^n\lambda^{-(2/3)(c/2 -1/4)n^2}
\sum_{k=1+n/2}^{\infty}
\lambda^{-(2/3)(1+2c)k^2}g_{\frac{n}{2}+k}g_{-\frac{n}{2}+k},
\label{Zn}
\end{align}
where the summations with respect to $k$ run over integers
(half-an-odd integers) when $n$ is even (odd).
Since the summation in $Y_n$ is convergent in the limit $n\to\infty$,
we have
\be
Y_n\sim \lambda^{-(2/3)(c/2 -1/4)n^2},
\quad(n\gg 1).
\ee
On the other hand, the large $n$ behavior of $Z_n$ is
\be
Z_n\sim \lambda^{-(2/3)(c/2 -1/4)n^2}\times
\lambda^{-(2/3)(1+2c)(1+n/2)^2}
\sim
\lambda^{-(2/3)cn^2},
\quad(n\gg 1).
\ee
Therefore, we have
\be
Y_n \gg Z_n \sim t_n,
\quad(n\gg 1),
\ee
and we can neglect $2 t_0t_n$ and $Z_n$ in the second term of
\eqref{eq_tn2} for a large $n$.
The constant $c$ is determined by the requirement that the first
term $(n^2-1)t_n\sim \lambda^{-(2/3)cn^2}$ of \eqref{eq_tn2} and the
leading part $\lambda^{1-(2/3)n^2}Y_n\sim \lambda^{-(c/3+1/2)n^2}$ of
the second term have the same large $n$ behavior:
\be
c=\frac32.
\ee
Therefore, the equation of motion \eqref{eq_tn2} for $t_n$ is reduced
to
\be
g_n=
\frac{1}{n^2-1}\sum_{k=1-n/2}^{n/2-1}\lambda^{-(8/3)k^2}
g_{\frac{n}{2}-k}g_{\frac{n}{2}+k}.
\label{eq_gn}
\ee
Although \eqref{eq_gn} has been derived for a sufficiently large $n$,
let us assume that it is valid for all $n (\ge 2)$.
Then, since the RHS of \eqref{eq_gn} is given in terms of
$g_k$ with $k$ smaller than $n$,
\eqref{eq_gn} determines $g_n$ recursively once we fix
$g_1=(\lambda^2/2)t_1$.
In fact, numerical analysis shows that \eqref{eq_gn} reproduces the
peculiar subleading behavior of fig.\ \ref{subleading}.

For analytic evaluation of $g_n$, let us make a further simplification
on \eqref{eq_gn} which can still reproduce the behavior of
$t_n$  observed in sec.\ 3.
This simplification is to keep only the term with the smallest
$\abs{k}$ in the summation of \eqref{eq_gn} by regarding
that $\lambda^{-8/3}=0.12$ is sufficiently small:
\be
g_n=\frac{1}{n^2-1}\times\begin{cases}
\ds \left(g_{n/2}\right)^2 &(n:{\rm even})\\[5pt]
\ds \eta\,g_{(n-1)/2}\,g_{(n+1)/2}& (n:{\rm odd})
\end{cases},
\label{eq_simple}
\ee
with
\be
\eta=2\lambda^{-2/3}=\frac{32}{27}.
\label{eta}
\ee

Eq.\ \eqref{eq_simple} is still too complicated to be solved
analytically for a generic $n$. However, we can obtain the solution
$g_n$ for specific values of $n$;
$n=2^m+2^{m-1}+\cdots +2^{m-a}=2^m(2-2^{-a})$ ($0\leq a\leq m$)
and $n=2^m+2^{m-b}=2^m(1+2^{-b})$ ($0\le b\le m$).
These $n$ are of the form $11\cdots 100\cdots0$ and
$10\cdots 010\cdots0$ in the binary notation.
For $n=2^m(2-2^{-a})$, \eqref{eq_simple} gives
\begin{align}
g_{2^{m+1}-2^{m-a}}
&=\frac{1}{(2^{m+1}-2^{m-a})^2-1}\left(g_{2^m-2^{m-a-1}}\right)^2  \nn\\
&=
\prod_{k=0}^{m-a-1}\biggr(\frac{1}{(2^{m+1-k}-2^{m-a-k})^2
 -1}\biggr)^{2^k}
\bigr(g_{2^{a+1}-1}\bigr)^{2^{m-a}},
\label{11a}
\end{align}
which in the special case of $a=0$ is the equation for $g_{2^m}$.
Then, using again \eqref{eq_simple}, $g_{2^{a+1}-1}$ on the RHS of
\eqref{11a} is given as
\be
g_{2^{a+1}-1}
=\frac{\eta}{(2^{a+1}-1)^2-1}g_{2^a}g_{2^a-1}
=\prod_{k=0}^{a-1}\frac{\eta}{(2^{a-k+1}-1)^2-1}
\prod_{\ell=0}^a g_{2^{a-\ell}},
\label{g2^(a+1)-1}
\ee
and the final factor $g_{2^{a\ell}}$ in \eqref{g2^(a+1)-1} is
expressed in terms of $g_1$
using \eqref{11a} with $a=0$ and $m=a-\ell$.
Completing the calculations we obtain $g_{n=2^m(2-2^{-a})}$ in a
closed form as a function of $g_1$ (see appendix \ref{appG}).
In particular, for a large $n=2^m(2-2^{-a})$ we have
\be
g_{n=2^m(2-2^{-a})}=
16\left(\frac{g_1}{G_a}\right)^n
\left(n^2-\frac{1}{7}+{\cal O}(n^{-2})\right).
\label{gna}
\ee
In \eqref{gna}, $G_a$ depends on only $a$ and is given by
\begin{align}
G_a&=\Biggl(\frac{2^{(a^2+5a+8)/2}}{\eta^a}\,
T_a^2\, e^{-S_a}
\prod_{k=1}^a(2^k-1)\cdot
\prod_{p=0}^{a-1}\prod_{q=0}^{a-p-1}
\left(2^{2(a-p-q)}-1\right)^{2^q}
\Biggr)^{1/T_a},
\label{G}
\end{align}
where $T_a$ and $S_a$ are defined by
\begin{align}
&T_a= 2^{a+1}-1 ,
\label{Ta}
\\
&S_a= \sum_{p=1}^{\infty}\frac{\left(2^{a+1}-1\right)^{-2p}}{
p\left(2^{2p+1}-1\right)}.
\label{Sa}
\end{align}
Similarly, for $n=2^m(1+2^{-b})$ we have
\be
g_{n=2^m(1+2^{-b})} = 16\left(\frac{g_1}{\wt{G}_b}\right)^n
\left(n^2 -\frac{1}{7}+{\cal O}(n^{-2})\right),
\label{gmb}
\ee
where $\wt{G}_b$ is given by
\be
\wt{G}_b=
\Biggl(
\frac{3\cdot 2^{(b^2+3b+8)/2}}{\eta^b}\wt{T}_b^2 e^{-\wt{S}_b}
\prod_{k=0}^{b-1}(2^k+1)
\cdot\prod_{p=0}^{b-2}
\prod_{q=0}^{b-p-2}\left(2^{2(b-p-q-1)}-1\right)^{2^q}
\Biggr)^{1/\wt{T}_b},
\label{wtG}
\ee
with
\begin{align}
&\wt{T}_b = 2^b+1,\\
&\wt{S}_b =\sum_{p=1}^{\infty}\frac{\left(2^b+1\right)^{-2p}}{
p\left(2^{2p+1}-1\right)}.
\end{align}
In \eqref{G} and \eqref{wtG}, the products $\prod_{k=n}^m$ with
$n>m$ are defined to be equal to one.
For example, the products in \eqref{G} are missing for $G_0$.

\begin{figure}[htbp]
\begin{center}
\leavevmode
\put(0,170){\boldmath{$G_a$,$\wt{G}_b$}}
\put(205,125){\boldmath{$n/2^m$}}
\epsfig{file=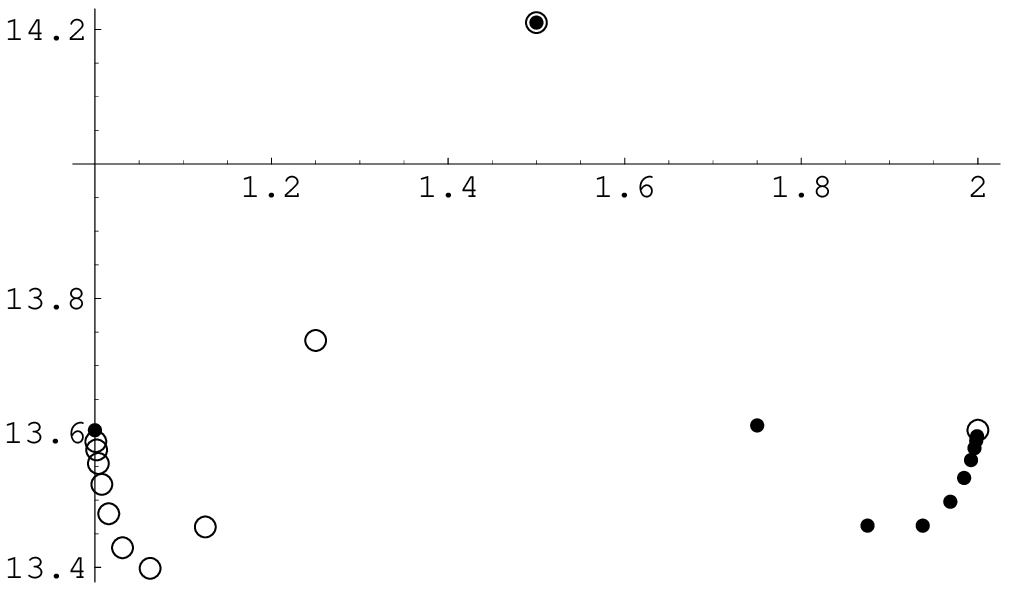, width=7cm}
\vspace{-1cm}
\caption{$G_a$ (dots) and $\wt{G}_b$ (circles). The horizontal axis is
$n/2^m=2-2^{-a}$ for $G_a$ and $n/2^m=1+2^{-b}$ for $\wt{G}_b$.
Here we use the original value $\eta=2\lambda^{-2/3}=32/27$.}
\label{gagb_eta-2ep}
\end{center}
\end{figure}

\begin{figure}
\begin{center}
\leavevmode
\put(0,170){\boldmath{$G_a$,$\wt{G}_b$}}
\put(205,75){\boldmath{$n/2^m$}}
\epsfig{file=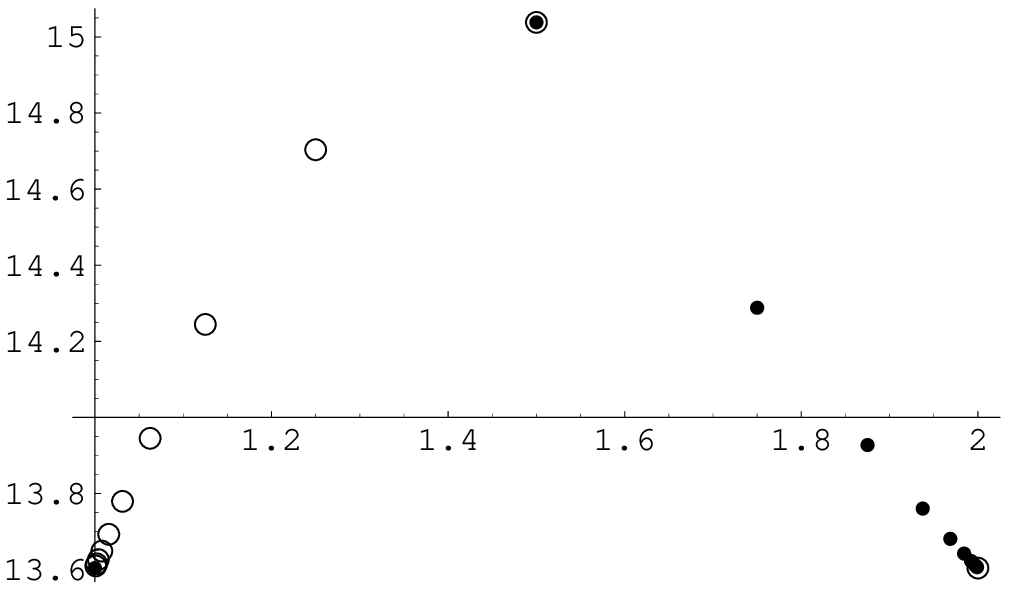, width=7cm}
\vspace{-1cm}
\caption{$G_a$ (dots) and $\wt{G}_b$ (circles) as functions of $n/2^m$
calculated by adopting $\eta=1$.}
\label{gagb_eta-1}
\end{center}
\end{figure}

Let us see the behaviors of $G_a$ and $\wt{G}_b$.
First, we have
\be
G_0=G_\infty=\wt{G}_0=\wt{G}_\infty=16\, e^{-S_0}=13.604 .
\ee
The equality of these four quantities is natural since $a=0$,
$a=\infty$, $b=0$ and $b=\infty$ all correspond to the points of the
same type $n=2^m$. The value $16\,e^{-S_0}$ is independent of $\eta$.
Next, figs.\ \ref{gagb_eta-2ep} and \ref{gagb_eta-1} show
$G_a$ (dots) and $\wt{G}_b$ (circles).
The horizontal axis is $2-2^{-a}$ for $G_a$ and
$1+2^{-b}$ for $\wt{G}_b$ (recall that $n=2^m(2-2^{-a})$ for $G_a$,
and $n=2^m(1+2^{-b})$ for $\wt{G}_b$).
Since $a=1$ and $b=1$ correspond to the same $n=2^m+2^{m-1}$, we have
$G_1=\wt{G}_1$.
The difference between figs.\ \ref{gagb_eta-2ep} and \ref{gagb_eta-1}
is that we have used the original value $\eta=32/27$ of \eqref{eta} in
the former,
while the smaller value $\eta=1$ has been adopted in the latter (see
below for the reason why we consider $\eta$ different from
\eqref{eta}).


Substituting \eqref{gna} and $g_1=(\lambda^2/2)t_1$
into \eqref{ansatz}, we get for $n\gg 1$
\be
t_{n=2^m(2-2^{-a})}
=-\frac{32}{\lambda}\biggr(-\frac{\lambda^2 }{2 G_a}\,t_1\biggr)^n
\lambda^{-n^2}\left(n^2-\frac{1}{7}+{\cal O}(n^{-2})\right).
\label{tn=32}
\ee
Similarly from \eqref{gmb}, we obtain $t_{n=2^m(1+2^{-b})}$, which is
given by \eqref{tn=32} with $G_a$ replaced by $\wt{G}_b$.
Let us compare this expression of $t_n$ with the results of our
numerical analysis of sec.\ 3.1. There $t_n$ was given as (we have put
$\xi=1$)
\be
t_n=-\subt_n\,(-\beta)^n\,\lambda^{-n^2}.
\label{tn=htn}
\ee
In sec.\ 3.1, we first determined $\beta$ (and $\xi$) by fitting $t_n$
using \eqref{tn=htn} with $\subt_n$ treated as $n$-independent
quantity $A$. We found that the subleading part $\subt_n$ has a
peculiar $n$-dependence of fig.\ \ref{subleading}.
The correspondence between \eqref{tn=32} and \eqref{tn=htn} should
be as follows:
\begin{align}
\beta(t_1)&=\frac{\lambda^2}{2\overline{G}}\,t_1,
\label{betaana}
\\
\subt_n &=\frac{32}{\lambda}
\left(n^2-\frac{1}{7}+{\cal O}(n^{-2})\right)
\biggr(\frac{\overline{G}}{G_a}\biggr)^n,
\label{subtana}
\end{align}
where $\overline{G}$ is the ``average value'' of $G_a$.
Here we can treat only the special points of the form
$n=2^m(2-2^{-a})$ and $2^m(1+2^{-b})$. However, $\overline{G}$ should
be regarded as an average of $G_a$ and $\wt{G}_b$ over all $n$ points
assuming that the formula like \eqref{tn=32} holds for a generic $n$.

Let us consider $\beta(t_1)$ of \eqref{betaana}.
If we take $\overline{G}=14$, which is a reasonable value as seen from
fig.\ \ref{gagb_eta-2ep}, we have $\lambda^2/(2\overline{G})=0.17$.
This is close to the proportionality constant of $\beta(t_1)$
determined numerically in sec.\ 3.1 (see below \eqref{betaproptot1}).

Eq.\ \eqref{subtana} cannot reproduce the peak
behavior of fig.\ \ref{subleading} if we adopt as $\eta$ the original
value \eqref{eta}, since $G_a$ and $\wt{G}_b$ has minima at
points a little off the points $n=2^m$ as seen from fig.\ \ref{gagb_eta-2ep}.
However, if we take a smaller value $\eta=1$ as in fig.\ \ref{gagb_eta-1},
\eqref{subtana} does reproduce the desired peaks at $n=2^m$
since, in particular,  $G_a$ and $\wt{G}_b$ take the smallest value at
these points. The critical value of $\eta$ below which $G_a$ and
$\wt{G}_b$ have minimum at $n=2^m$ is $1.036$.
One would think that adopting $\eta$ different from the original
value \eqref{eta} is groundless.
However, there is no reason to stick to the original value \eqref{eta}
in solving \eqref{eq_simple}
since we have made two steps of approximations (the first is
the large $n$ approximation and the second is that of
$\lambda^{-8/3}\ll 1$) in reaching \eqref{eq_simple} from the original
equation of motion.
It is expected that, by incorporating the effects in making the
approximations, the ``effective value'' of $\eta$ in \eqref{eq_simple}
is reduced from the original one.

\section{Summary and discussions}

In this paper, we studied time dependent and spatially homogeneous
solutions of the truncated version of cubic string field theory.
We expanded the component fields in terms of $\cosh nx^0$ modes, wrote
down the equations of motion of the modes, and solved them both
numerically and analytically.
Our finding in this paper is that the tachyon modes have the leading
$n$-dependence of \eqref{tnfit} multiplied by the subleading
dependence with peaks at $n=2^m=4,8,16,32,64,128,\cdots$.

Let us return to the question posed in sec.\ 1: what the tachyon
profile would be if we carry out summations over all modes.
First, assuming that $t_n$ is simply given by its leading part
\eqref{tnfit} with $\xi=1$ for all $n (\ge 1)$, the profile $t(x^0)$
is given by
\be
t(x^0)=t_0-A\sum_{n=1}^\infty\lambda^{-n^2}(-\beta)^n
\cosh nx^0 .
\label{sum1}
\ee
Unfortunately,
this function does not have a desirable behavior as a rolling
tachyon solution. It is an oscillating function of $x^0$ which grows
like $\exp\left((x^0)^2/(4\ln\lambda)\right)$ for large
$x^0$.\footnote{This is seen from the formula
$$
\sum_{n=-\infty}^\infty(-1)^n e^{-an^2 -b|n|}\,e^{nx}
=e^{x^2/(4a)}\sum_{m=-\infty}^\infty\left(
e^{-2|m|b}\,e^{-(x-4ma)^2/(4a)}
-e^{-|2m-1|b}\,e^{-(x-2(2m-1)a)^2/(4a)}\right)
$$
with $a$ and $b$ identified with $\ln\lambda$ and $\ln(1/\beta)$,
respectively.
}

The function \eqref{sum1}, however, has other apparently equivalent
expressions. Rewriting $\lambda^{-n^2}$ in \eqref{sum1} into
$\lambda^{-\p_0^2}$ and moving it outside the summations,
we obtain the second expression:
\be
t(x^0)=t_0+A-\frac{A}{2}\lambda^{-\p_0^2}\left(
\frac{1}{1+\beta\,e^{x^0}}+\frac{1}{1+\beta\,e^{-x^0}}\right),
\label{sum2}
\ee
with $\lambda^{-\p_0^2}$ defined by
$\lim_{M\to\infty}\sum_{k=0}^M (-\ln\lambda)^k/k!\cdot \p_0^{2k}$.
However, the limit $M\to\infty$ does not seem to exist since,
as $M$ becomes larger, \eqref{sum2} becomes rapidly oscillating with
growing amplitude for intermediate values $x^0\sim\ln(1/\beta)$.

\begin{figure}
\begin{center}
\leavevmode
\put(207,152){\boldmath{$x^0$}}
\epsfig{file=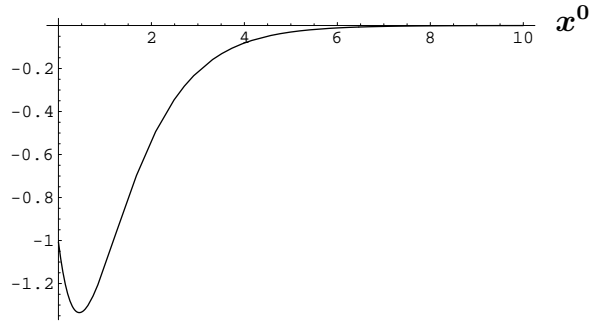, width=7cm}
\vspace{-1cm}
\caption{$-\sum_{n=1}^{\infty}\lambda^{-n^2}(-1)^n
\left(\beta^n-\beta^{-n}\right)e^{-nx}$
at $\beta =0.1$. }
\label{exp3}
\end{center}
\end{figure}

The third definition of $t(x^0)$ is obtained by reexpanding the
quantity in the parenthesis of \eqref{sum2} in terms of the power
series of $e^{-x^0}$ and then moving $\lambda^{-\p_0^2}$ inside the
series:
\be
t(x^0)=t_0+\frac{A}{2}-\frac{A}{2}\sum_{n=1}^\infty
\lambda^{-n^2}(-1)^n\left(\beta^n-\beta^{-n}\right)e^{-nx^0}.
\label{sum3}
\ee
This function has a convergent limit as $x^0\to\infty$.
However, it is not a monotonic function of $x^0$, and what is worse,
its derivative at $x^0=0$ is not equal to zero in spite of the fact
that we started with $\cosh nx^0$ modes (see fig.\ \ref{exp3}).

The origin of the phenomenon that apparently equivalent expressions of
$t(x^0)$ give completely different functions is the fact that the
insertion of $\lambda^{-n^2}$ makes any series with finite radius of
convergence into another series with infinite radius of convergence.
This problem is related to the long standing problem of how to treat
the infinite derivative operators like  $\lambda^{-\p_0^2}$ in string
field theories \cite{HataNL,EliWoo}.

The above argument shows that the leading part \eqref{tnfit} alone
cannot give desirable tachyon profiles.
Inclusion of the subleading part $\subt_n$ \eqref{subt_n} could
lead to $t(x^0)$ with a desirable profile.
For studying this possibility, we have to solve the equation of motion
to obtain analytically the  subleading part $\subt_n$ for all $n$ and
carry out the summation over $n$.
However, it is very likely that the problem of $\lambda^{-\p_0^2}$
persists even if we take into account the subleading part.
Another possible way of getting a desirable tachyon profile would be
to claim that the tachyon profile should be given by
$T(x^0)\equiv \lambda^{\p_0^2}t(x^0)$ rather than $t(x^0)$ itself:
this new $T(x)$ is equal to $t(x)$ in the translationally invariant
case, and it gives a monotonic and convergent profile for $t(x^0)$ of
\eqref{sum2}.
Besides the tachyon profile itself, we have to study the time
dependence of other physical quantities such as the energy-momentum
tensor \cite{padic,Yang}.

In this paper we have studied time-dependent solutions in truncated
CSFT. It is interesting to construct time-dependent exact solutions of
the full CSFT \cite{kluson}.
The tachyon vacuum solution of \cite{takahashi} would be a
staring point of the construction.
Study of time dependent solutions in vacuum string field theory
\cite{RSZ,HataVSFT,Okawa} is also an interesting research subject.

\noindent
\section*{Acknowledgments}

We would like to thank S.\ Sugimoto, T.\ Takahashi and I.\ Kishimoto
for valuable discussions.
The work of H.\,H. was supported in part by a
Grant-in-Aid for Scientific Research from Ministry of Education,
Culture, Sports, Science, and Technology (\#12640264).

\appendix
\section{$\bm{g_n}$ for specific values of $\bm{n}$}
\label{appG}

In this appendix we complete solving \eqref{eq_simple} for $g_n$ at
$n=2^m(2-2^{-a})$  and deriving the expression \eqref{gna}.
The derivation of $g_n$ at $n=2^m(1+2^{-b})$ is quite similar.

For $g_{n=2^m(2-2^{-a})}$ we have \eqref{11a}, \eqref{g2^(a+1)-1}
and
\be
g_{2^{a-\ell}}
=\prod_{k=0}^{a-\ell-1}\left[\frac{1}{(2^{a-\ell-k})^2-1}\right]^{2^k}
(g_1)^{2^{a-\ell}}.
\label{calc_2}
\ee
Substituting \eqref{g2^(a+1)-1} and \eqref{calc_2} into
\eqref{11a} we obtain a closed expression for $g_{n=2^m(2-2^{-a})}$:
\begin{align}
g_{n=2^m(2-2^{-a})}&=
\prod_{k=0}^{m-a-1}\biggr(\frac{1}{(2^{m+1-k}-2^{m-a-k})^2 -1}
\biggr)^{2^k}\cdot
\biggr(\prod_{q=0}^{a-1}\frac{\eta}{(2^{a-q+1}-1)^2-1}\biggr)^{2^{m-a}}
\nn\\
&\hspace*{2cm}\times
\biggr(\prod_{\ell=0}^a
\prod_{p=0}^{a-\ell-1}\left[\frac{1}{(2^{a-\ell-p})^2-1}\right]^{2^p}
\biggr)^{2^{m-a}}
\left(g_1\right)^n,
\label{gnany}
\end{align}
which is valid for any $m$ and $a$ ($0\le a\le m$).
For obtaining the approximate expression \eqref{gna}, we rewrite the
first product in \eqref{gnany} into
\be
\prod_{k=0}^{m-a-1}
\biggr(\frac{1}{(2^{m+1-k}-2^{m-a-k})^2 -1}\biggr)^{2^k}
=\prod_{k=0}^{m-a-1}\left(n^2\,2^{-2k}\right)^{-2^k}
\prod_{\ell=0}^{m-a-1}\left(1-n^{-2} 2^{2\ell}\right)^{-2^\ell}.
\label{dodemo1}
\ee
The first product on the RHS of \eqref{dodemo1} is given by
\be
\prod_{k=0}^{m-a-1}\left(n^2\,2^{-2k}\right)^{-2^k}
=2^4 n^2\left[(4 T_a)^{2/T_a}\right]^{-n},
\label{dodemo2}
\ee
and the second product is evaluated for a large $n$ as follows:
\begin{align}
&\prod_{\ell=0}^{m-a-1}\left(1-n^{-2} 2^{2\ell}\right)^{-2^\ell}
=\exp\left(-\sum_{\ell=0}^{m-a-1}2^{\ell}
\ln\left(1-n^{-2} 2^{2\ell}\right)\right)
\nn\\
&=\exp\left(\sum_{p=1}^\infty\frac{n^{-2p}}{p}
\sum_{\ell=0}^{m-a-1}2^{(1+2p)\ell}\right)
=\exp\left(\sum_{p=1}^\infty\frac{n\,T_a^{-2p-1}-n^{-2p}}{
p(2^{2p+1}-1)}\right)
\nn\\
&=\exp\left(\frac{S_a}{T_a}\,n -\frac17\,n^{-2}+
\calO(n^{-4})\right).
\label{dodemo3}
\end{align}
Plugging \eqref{dodemo1} with \eqref{dodemo2} and \eqref{dodemo3} into
\eqref{gnany} and rearranging other factors in \eqref{gnany}, we get
\eqref{gna}.

\end{document}